\documentclass[dvipdmfx]{PoS}
\usepackage{amsmath}
\usepackage{url}
\usepackage{here}
\usepackage{color}

\makeatletter
\newcommand{\figcaption}[1]{\def\@captype{figure}\caption{#1}}
\newcommand{\tblcaption}[1]{\def\@captype{table}\caption{#1}}
\newcommand{\MSbar}{\overline{\mathrm{MS}}}
\newcommand{\TGF}{\mathrm{TGF}}
\newcommand{\SF}{\mathrm{SF}}
\newcommand{\Lmax}{L_{\mathrm{max}}}
\newcommand{\figwidth}{0.66}
\makeatother

\title{Numerical determination of the $\Lambda$-parameter in SU(3) gauge theory from the twisted gradient flow coupling}

\ShortTitle{Numerical determination of the $\Lambda$-parameter in SU(3) gauge theory from the twisted gradient flow coupling}

\author{	Ken-Ichi Ishikawa$^{a,b}$, 
			Issaku Kanamori$^{a}$, 
			Yuko Murakami$^{a}$, 
			Ayaka Nakamura$^{a}$, 
			Masanori Okawa$^{a,b}$ and 
			\speaker{Ryoichiro Ueno$^{a}$}
			\\
			{$^{a}$}Graduate School of Science, Hiroshima University, Higashi-Hiroshima, Hiroshima 739-8526, Japan
			\\
			{$^{b}$}Core of Research for the Energetic Universe, Hiroshima University, Higashi-Hiroshima, Hiroshima 739-8526, Japan
			\\
			E-mail:\email{ryoichiro-ueno@hiroshima-u.ac.jp}}

\abstract{We estimate the $\Lambda$-parameter in the $\MSbar$ scheme for the SU(3) pure gauge theory with the twisted gradient flow method non-perturbatively. We obtain  $\Lambda_{\MSbar}/\sqrt{\sigma}=0.527(13)(10)$ and $r_{0}\Lambda_{\MSbar}=0.605(15)(5)$ which are consistent with the known values.This demonstrates the validity of the present method.
{\normalsize\vspace*{-33.5em}\begin{flushright}\ \ \ \ HUPD-1614\end{flushright}\vspace*{31em}\ }}

\FullConference{34th annual International Symposium on Lattice Field Theory\\ 24-30 July 2016\\ University of Southampton, UK}

\begin{document}


\section{Introduction}
\label{sec:introduction}

$\Lambda$-parameter is a fundamental scale parameter in asymptotic free gauge theories. The non-perturbative determination of the $\Lambda$-parameter in QCD has its phenomenological importance, and a huge amount of effort using lattice QCD has been made to the determination. In this study, we numerically evaluate the $\Lambda$-parameter in the $\MSbar$ scheme for the SU(3) pure gauge theory by lattice simulations non-perturbatively using the twisted gradient flow (TGF) scheme recently proposed by Ramos \cite{Ramos:TGF}.

The gradient flow scheme is one of the application of the gradient flow method, in which the gauge field is smeared with the so-called flow equation and the smeared gauge field has a nice perturbative property on the renormalizability \cite{Narayanan:2006rf,Luscher:WilsonFlow,Luscher/Weisz:GF}. Ramos has investigated the TGF coupling for the SU(2) pure Yang-Mills theory \cite{Ramos:TGF}. We extend his study to the SU(3) pure Yang-Mills theory.  In addition to this, we extract the $\Lambda$-parameter in the TGF scheme and convert it to the $\MSbar$ scheme. This study could be an entirely self-consistent determination of $\Lambda_{\MSbar}$ for the SU(3) pure gauge theory with the TGF scheme. Various coupling schemes defined via the gradient flow method have been proposed and investigated in Refs. \cite{Fodor/Holland/Kuti/Nogradi:pGF,Lin:2015zpa,Fritzsch/Ramos:GF,Brida:2015gqj,Leino:2015bfg}.

The one-loop perturbative relation between the TGF coupling and the $\MSbar$ coupling is required to obtain $\Lambda_{\MSbar}$ from $\Lambda_{\TGF}$. This is not yet known in the literature, and an ongoing study on the matching between the $\MSbar$ and $\TGF$ schemes is presented by E. Ibanez Bribian and M. Garcia Perez in this conference \cite{bribian}. In this study we employ the Schr\"{o}dinger functional (SF) scheme \cite{SFREFS,Sint/Sommer:LambdaSF}, one of the finite size box scheme, as the intermediate scheme to bypassing the direct conversion from the $\TGF$ scheme to the $\MSbar$ scheme. We numerically evaluate the one-loop relation between the TGF coupling and the SF coupling by lattice simulations in the weak coupling region to have the ratio $\Lambda_{\SF}/\Lambda_{\TGF}$. Combined with the known ratio $\Lambda_{\SF}/\Lambda_{\MSbar}$ \cite{Sint/Sommer:LambdaSF}, we can obtain $\Lambda_{\MSbar}/{\Lambda_{\TGF}}$.

Our strategy to obtain $\Lambda_{\MSbar}/A_{\mathrm{phys}}$ is summarized as follows:
\begin{align}
	\frac{\Lambda_{\MSbar}}{A_{\mathrm{phys}}}=\frac{\Lambda_{\MSbar}}{\Lambda_{\SF}}\cdot\frac{\Lambda_{\SF}}{\Lambda_{\TGF}}\cdot\frac{\Lmax\Lambda_{\TGF}}{\Lmax A_{\mathrm{phys}}},
	\label{eq:strategy}
\end{align}
where $A_{\mathrm{phys}}$ is a physical mass scale defined through a low energy (hadronic scale) observable, and $\Lmax$ is a maximum box size at which the TGF coupling is renormalized. $\Lmax$ is a reference scale and chosen so that we can make contact with the low-energy scale $A_{\mathrm{phys}}$ using the renormalized coupling constant. We employ the string tension $\sqrt{\sigma}$ or the Sommer scale $r_0$ as the low energy observables. The high precision lattice data for $\sqrt{\sigma}$ and $r_0$ are taken from Refs. \cite{Allton/Teper/Trivini:StringTension,Antonio/Okawa:StringTension} and \cite{Necco:Dthesis} respectively.

This paper is organized as follows. In section \ref{sec:setup}, we explain our simulation setup for the calculation of the TGF coupling. The numerical results for the step scaling of the TGF coupling, the low energy observables in $\Lmax$ unit, and the ratio $\Lambda_{\SF}/\Lambda_{\TGF}$ are presented in the following sections. Combining all pieces obtained, we give the preliminary result of $\Lambda_{\MSbar}/A_{\mathrm{phys}}$ in the last section.


\section{Simulation setup}
\label{sec:setup}

We employ the SU(3) Wilson gauge action in a box of size $L^4$ with the twisted boundary condition in the $x$--$y$ plane and periodic in the $z$--$t$ plane. For the details of the definition of the TGF coupling, the gradient flow equation, and the twisted boundary condition, we follow Ref. \cite{Ramos:TGF}. We employ the clover leaf definition for the field strength used in the TGF coupling. The renormalization scale $\mu=1/(cL)=1/\sqrt{8t}$ for the TGF couping is defined through the gradient flow time $t$ and the finite box size $L$. In this study we set $c=0.3$, which defines the scheme, for the renormalization scale.

We generate the gauge configurations using the heat-bath algorithm and measure the TGF coupling. In order to compute the TGF coupling we take five values for the lattice size: $L/a=12$, 16, 18, 24 and 32. Several values of the bare coupling $\beta=6/g_{0}^{2}$ are taken from the range $\beta\in [6.1, \cdots ,10.0]$ for each lattice size.


\section{TGF coupling and $\Lambda_{\TGF}$}
\label{sec:TGF-Lambda}

The discrete beta function at a finite cutoff ``$a$'' is defined by
\begin{align}
	B_{s}(u=g_{\TGF}^{2},a/L)=\frac{g_{\TGF}^{2}(s[a/L],\beta)-g_{\TGF}^{2}(a/L,\beta)}{\log(s^{2})},
	\label{lat-beta}
\end{align}
where $g_{\TGF}^{2}(a/L,\beta)$ is the TGF coupling measured at $\beta$ on a $(L/a)^4$ box. We use $s=3/2$ for the step scaling size in this work. 

The continuum limit of Eq. (\ref{lat-beta}) is obtained keeping the value of the renormalized coupling constant $g_{\TGF}^{2}(a/L,\beta)$ at a fixed value $u=g_{\TGF}^{2}(a/L,\beta)$. To do this we fit all the data evaluated at $L/a=12,16,18,24,32$ in $\beta=6.1$--$10.0$ with a polynomial function of $u$ and $a/L$. We obtain
\begin{align}
	& B_{3/2}(u,a/L)=\left[\sigma_{0}-1.76(54)(a/L)^{2}\right]u^{2}+\left[\sigma_{1}+2.83(40)(a/L)^{2}\right]u^{3}
	\nonumber\\
	&\quad+\left[0.000679(57)-0.81(10)(a/L)^{2}\right]u^{4}+\left[-0.0000608(91)+0.0608(93)(a/L)^{2}\right]u^{5},
	\label{eq:disc-beta}
\end{align}
where $\sigma_{0}=-b_{0}$ and $\sigma_{1}=\sigma_{0}^{2}\log[s^{2}]-b_{1}$ are the constants with the universal one/two-loop beta functions $b_{0/1}$. We use the fit ansatz incorporating the fact that the cut-off error is $O(a^2)$ in the pure gauge theory. The error in the numerics indicates the statistical error estimated from a random re-sampling method assuming a Gaussian distribution in the original data set.

\begin{figure}[t]
	\begin{tabular}{cc}
		\begin{minipage}{0.5\hsize}
			\centering
			\includegraphics[width=\figwidth\textwidth,angle=-90]{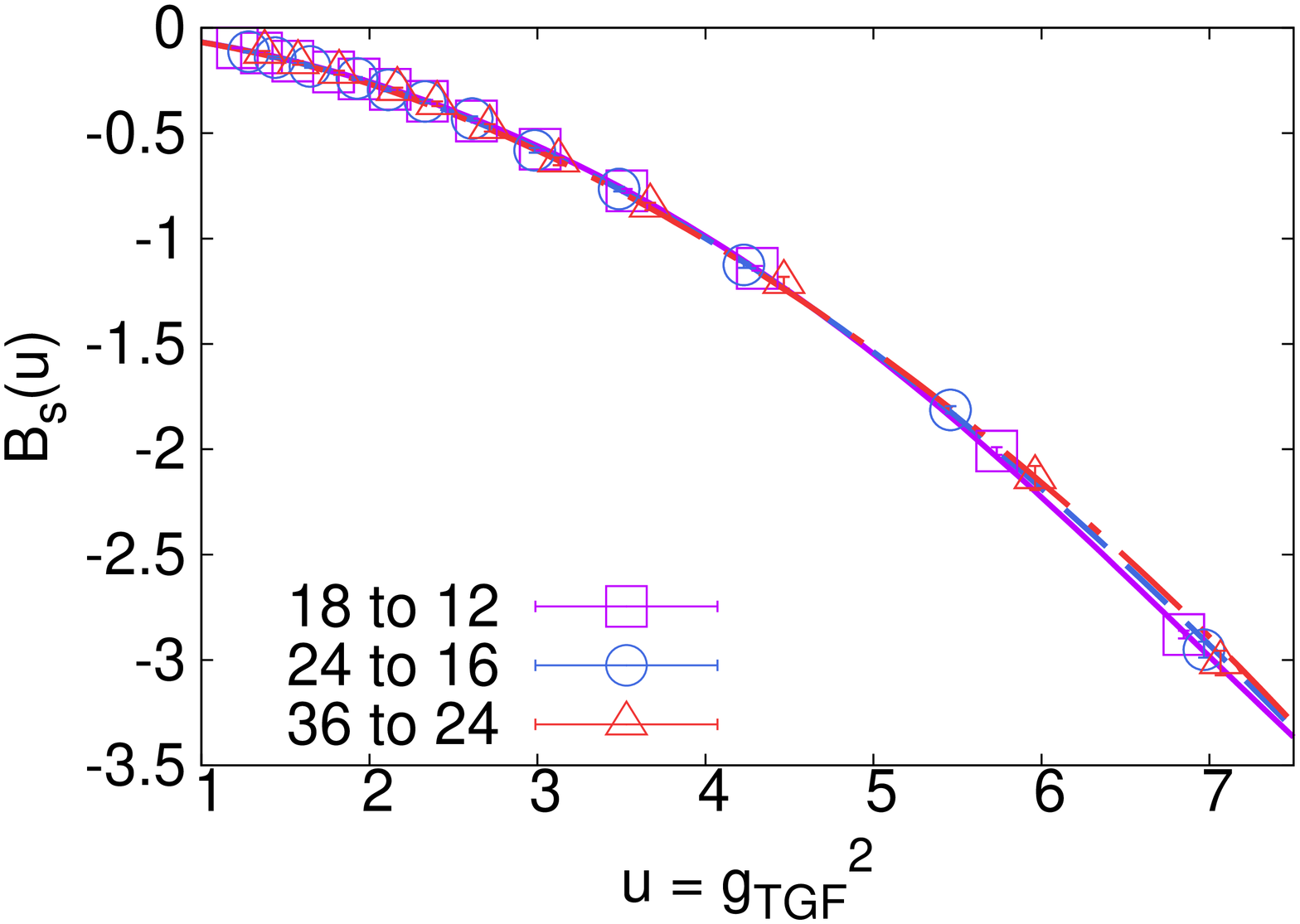}
		\end{minipage}
		\begin{minipage}{0.5\hsize}
			\centering
    	 	\includegraphics[width=\figwidth\textwidth,angle=-90]{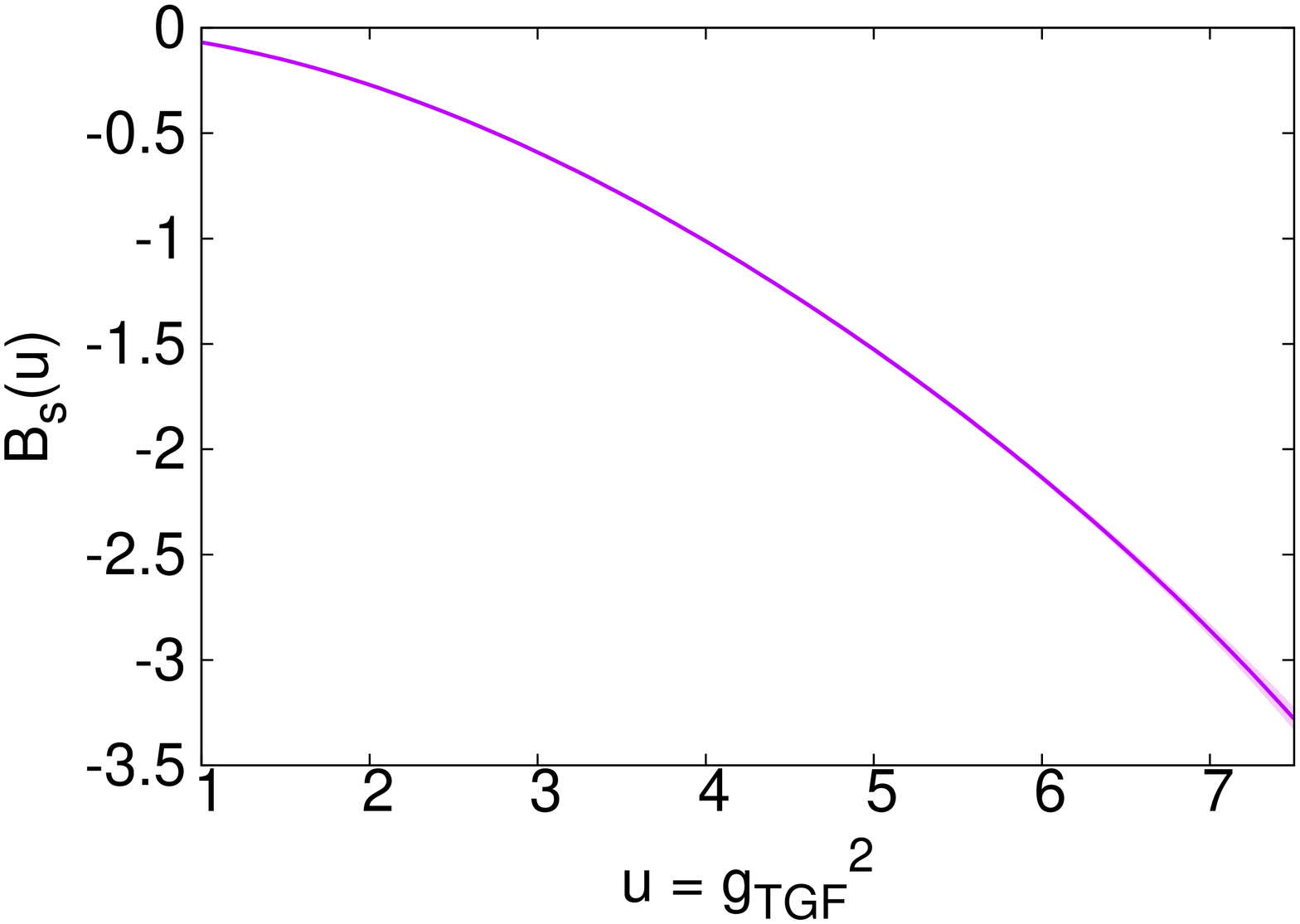}
		\end{minipage}
	\end{tabular}
	\caption{The discrete beta function for each lattice size (left) and in the continuum limit (right).}
	\label{fig:B}
\end{figure}

Figure \ref{fig:B} shows the discrete beta function $B_{s}(u,a/L)$ and the fit result. The fit yields $\chi^{2}/\mathrm{DoF}=0.96(46)$ indicating a good fitting. 

\begin{table}[t]
	\begin{tabular}{cc}
		\begin{minipage}[t]{.4\textwidth}
			\centering
			\begin{tabular}{cc}
				\hline\hline
				$g_{\TGF}^{2}(1/\Lmax)$ & $c\Lmax\Lambda_{\TGF}$
				\\ \hline
				6.0                     & 0.580(13)
				\\
				6.1                     & 0.589(13)
				\\
				6.2                     & 0.598(14)
				\\
				6.3                     & 0.606(14)
				\\
				6.4                     & 0.614(14)
				\\
				6.5                     & 0.622(14)
				\\
				6.6                     & 0.630(14)
				\\
				6.7                     & 0.638(15)
				\\
				6.8                     & 0.645(15)
				\\
				6.9                     & 0.652(15)
				\\
				7.0                     & 0.658(15)
				\\ \hline\hline
			\end{tabular}
			\caption{$\Lmax\Lambda_{\TGF}$.}
			\label{tab:LambdaTGF}
		\end{minipage}
		\hspace{2mm}\hfil
		\begin{minipage}[t]{0.6\hsize}
			\centering
			\begin{tabular}{ccc}
				\hline\hline
				$g_{\TGF}^{2}(1/\Lmax)$ & $\Lmax\sqrt{\sigma}$ & $\Lmax/r_{0}$
				\\ \hline
				6.0                     & 1.9244(79)           & 1.6980(86)
				\\
				6.1                     & 1.9546(76)           & 1.7188(85)
				\\
				6.2                     & 1.9816(77)           & 1.7415(88)
				\\
				6.3                     & 2.0022(77)           & 1.7593(87)
				\\
				6.4                     & 2.0368(76)           & 1.7772(86)
				\\
				6.5                     & 2.0588(76)           & 1.7969(85)
				\\
				6.6                     & 2.0858(77)           & 1.8172(84)
				\\
				6.7                     & 2.1093(76)           & 1.8343(85)
				\\
				6.8                     & 2.1367(77)           & 1.8483(85)
				\\
				6.9                     & 2.1587(77)           & 1.8645(84)
				\\
				7.0                     & 2.1818(79)           & 1.8821(84)
				\\ \hline\hline
			\end{tabular}
			\caption{$\Lmax\sqrt{\sigma}$ and  $\Lmax/r_{0}$.}
			\label{tab:string-tension.etc}
		\end{minipage}
	\end{tabular}
\end{table}

The RG evolution of the coupling can be traced using the discrete beta function, from which we can extracted the $\Lambda$-parameter. The $\Lambda$-parameter in the TGF scheme is approximated by
\begin{align}
	c\Lmax\Lambda_{\TGF}\simeq s^{n}\left(b_{0}g^{2}_{\TGF}(s^{n}/\Lmax)\right)^{-\frac{b_{1}}{2b_{0}^{2}}}\exp\left(-\frac{1}{2b_{0}g^{2}_{\TGF}(s^{n}/\Lmax)}\right),
	\label{eq:TGF-Lambda-scaled}
\end{align}
where $g_{\TGF}^{2}(s^{n}/\Lmax)$ is obtained after $n$-step RG evolution starting from $g_{\TGF}^{2}(1/\Lmax)$. The approximation becomes accurate when $g_{\TGF}^{2}(s^{n}/\Lmax)$ is sufficiently small. We use $n=200$ and Eq. (\ref{eq:TGF-Lambda-scaled}) can be used as the definition of the $\Lambda$-parameter.

In order to make contact with a low energy scale (hadronic scale), it is preferable to take the size $\Lmax$ to be as large as possible. This is equivalent to evolve $g^{2}_{\TGF}(1/\Lmax)$ from a larger value. We take several values for $g_{\TGF}^{2}(1/\Lmax)$ between 6.0 and 7.0 as the start point. Table \ref{tab:LambdaTGF} shows the resulting $\Lmax\Lambda_{\TGF}$.


\section{Physical scale, $\sqrt{\sigma}$ and $r_{0}$}
\label{sec:physical-scale}

To fix the physical mass scale $A_{\mathrm{phys}}$, we employ the string tension $\sqrt{\sigma}$ and the Sommer scale $r_0$. The mass scale $A_{\mathrm{phys}}$ must be counted by $\Lmax$ to relate them with $\Lmax\Lambda_{\TGF}$ obtained above.

We employ the data set of the string tension and the Sommer scale from Refs. \cite{Allton/Teper/Trivini:StringTension,Antonio/Okawa:StringTension} 
and \cite{Necco:Dthesis} respectively. These data are evaluated with the same action used in this study. We evaluate $(\Lmax/a)(aA_{\mathrm{phys}})$ at a fixed value of $u=g_{\TGF}^{2}(a/\Lmax,\beta)$ on several lattice sizes $\Lmax/a$ and $\beta$ values by harmonizing hadronic data and our coupling data. We extrapolate them into the continuum limit $a\to0$ with a linear function in $(a/\Lmax)^{2}$. The results for $\Lmax\sqrt{\sigma}$ and $\Lmax/r_{0}$ are listed in Table \ref{tab:string-tension.etc}.


\section{$\Lambda$-parameter ratio $\Lambda_{\SF}/\Lambda_{\TGF}$}
\label{sec:ratio}

\begin{figure}[h]
	\begin{minipage}{0.5\hsize}
		\centering
		\includegraphics[width=\figwidth\textwidth,angle=-90]{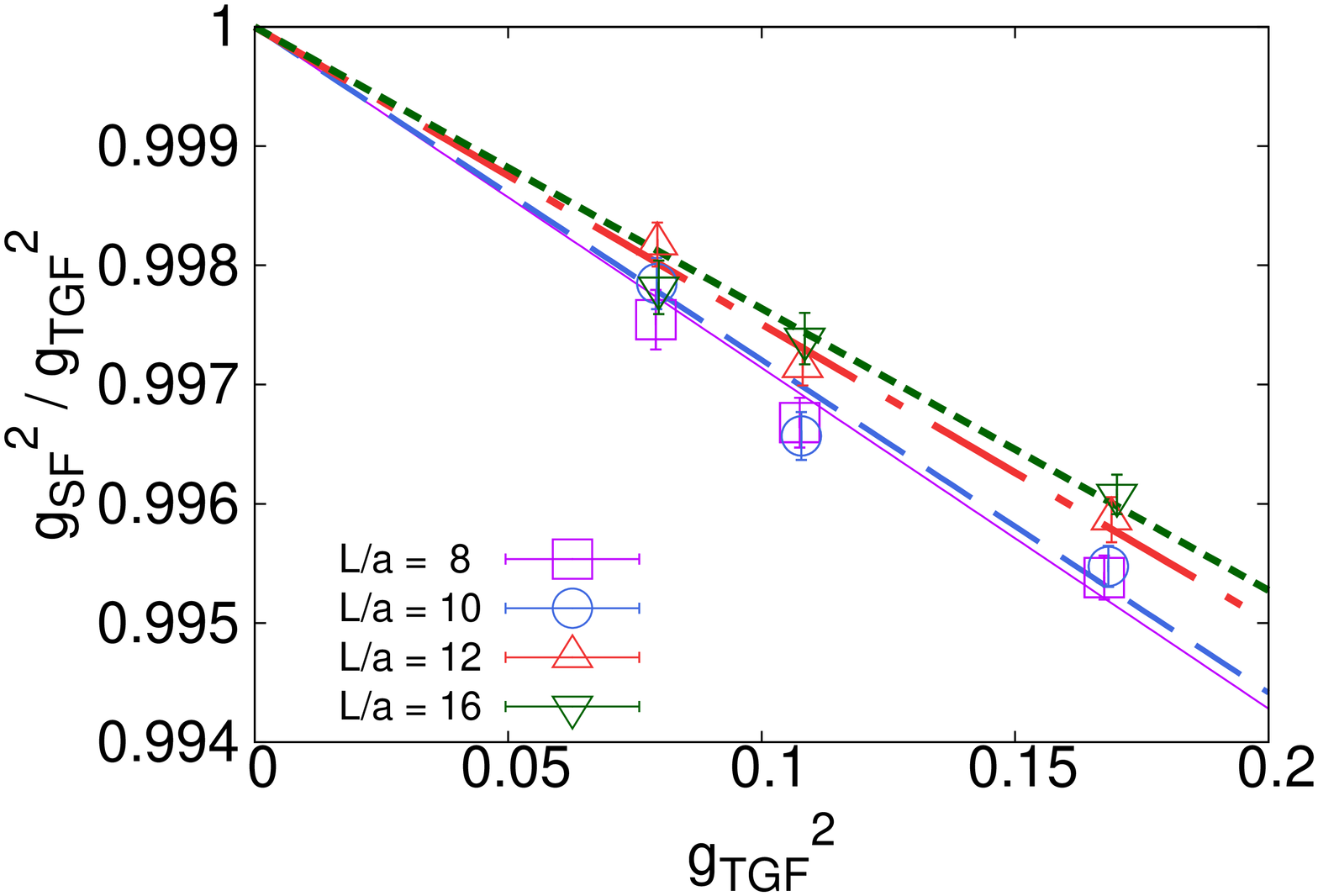}
		\caption{The ratio of the SF to the TGF couplings.}
		\label{fig:gSF/gTGF}
	\end{minipage}
	\begin{minipage}{0.5\hsize}
		\centering
		\includegraphics[width=\figwidth\textwidth,angle=-90]{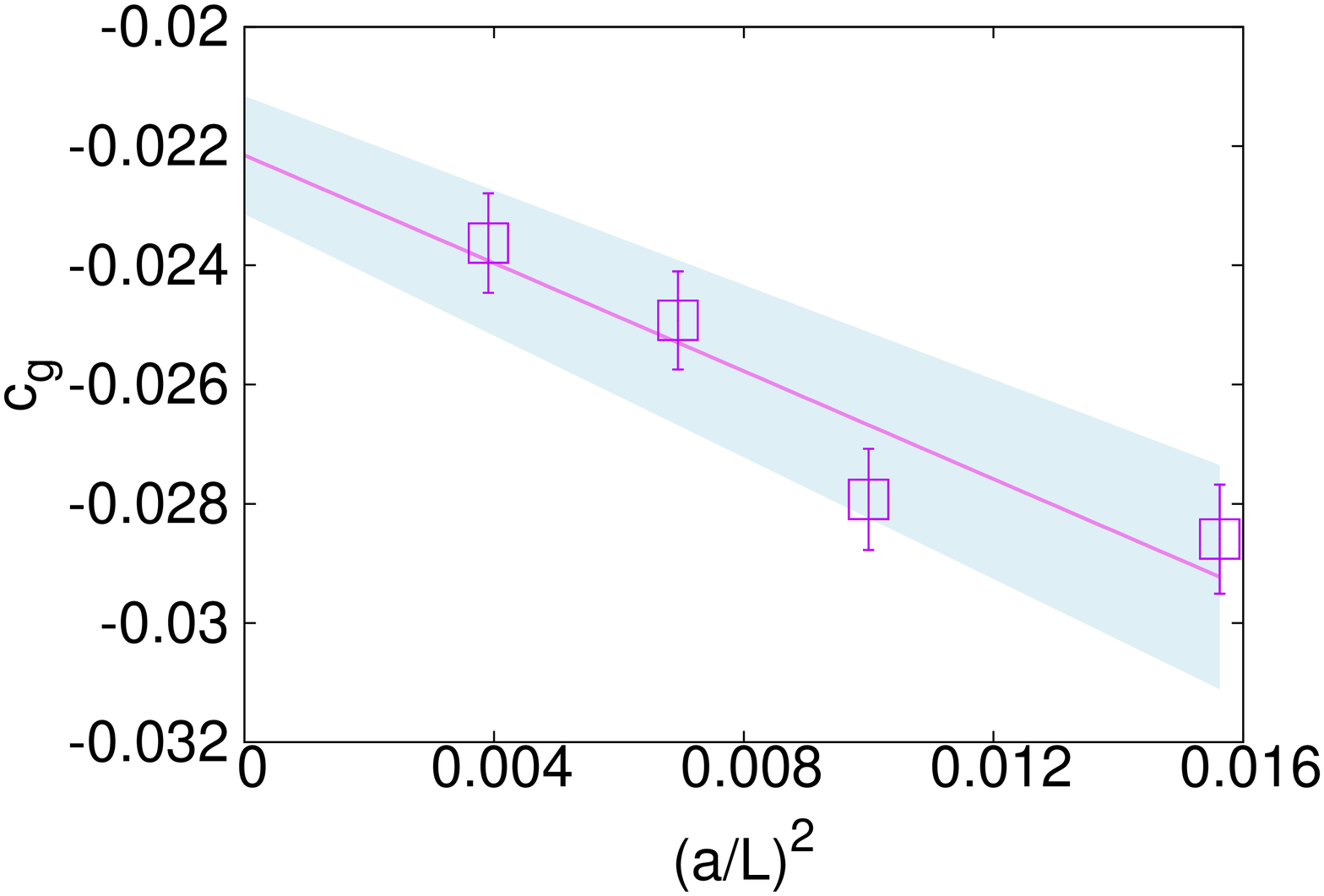}
		\caption{The continuum limit for $c_{\mathrm{g}}$.}
		\label{fig:c}
	\end{minipage}
\end{figure}

\begin{table}[t]
	\centering
	\begin{tabular}{ccc|ccc}
		\hline\hline
		$L/a$ & $c_{\mathrm{g}}(L/a)$ & $\chi^{2}/\mathrm{DoF}$ & $L/a$ & $c_{\mathrm{g}}(L/a)$ & $\chi^{2}/\mathrm{DoF}$
		\\ \hline
		 8    & $-$0.02859(92)        & 1.42                    & 12    & $-$0.02492(82)        & 0.98
		\\
		10    & $-$0.02793(85)        & 2.76                    & 16    & $-$0.02363(84)        & 1.11
		\\ 
		\hline\hline
	\end{tabular}
	\caption{The fit results for $c_{\mathrm{g}}$ at each lattice.}
	\label{tab:c}
\end{table}

The $\Lambda$-parameter ratio between the TGF scheme and the SF scheme is defined by
\begin{align}
	\frac{\Lambda_{\SF}}{c\Lambda_{\TGF}}&=\exp\left(\frac{c_{\mathrm{g}}^{(0)}}{2b_{0}}\right), 
	\label{eq:def-lambda-parameter-ratio}
\end{align}
where $c_{\mathrm{g}}^{(0)}$ is the one-loop coefficient in the SF coupling expanded by the TGF coupling. As the perturbative calculation is not yet available, we numerically estimate it in a weak coupling region on the lattice.

In order to compute the TGF coupling and the SF coupling, we take $L/a=8$, 10, 12 and 16 for the lattice size. The configurations are generated at $\beta=40$, 60 and 80 for each lattice size. The configurations with the SF boundary condition (including the boundary $O(a)$-improvement term) is independently generated with the same parameters ($\beta$ and $L/a$).

Thus we can evaluate the renormalized coupling in both schemes separately at the same bare couping and the lattice size using the same plaquette action with different boundary condition. Then the ratio $g_{\SF}^{2}(a/L,\beta)/g_{\TGF}^{2}(a/L,\beta)$ can be obtained and fitted as a function of $g_{\TGF}^{2}(a/L,\beta)$. Thus the one-loop coefficient can be extracted from
\begin{align}
	\frac{g_{\SF}^{2}(a/L,\beta)}{g_{\TGF}^{2}(a/L,\beta)}&=1+c_{\mathrm{g}}(a/L)g_{\TGF}^{2}(a/L,\beta)+\dots,
	\label{eq:coupling-ratio}\\
	c_{\mathrm{g}}(L/a)&=c_{\mathrm{g}}^{(0)}+c_{\mathrm{g}}^{(1)}(a/L)^{2}+\dots,
	\label{eq:c_g}
\end{align}
where we apply the $O(a)$-improvement in the SF scheme so that the cut-off error is $O(a^2)$.

We show $g_{\SF}^{2}/g_{\TGF}^{2}$ in Figure \ref{fig:gSF/gTGF}. Lines show the results from linear fitting. The fit results are summarized in Table \ref{tab:c}. Figure \ref{fig:c} shows the extrapolation to the continuum limit. Here $O(a^2)$-scaling is observed as expected. We obtain
\begin{align}
	 c_{\mathrm{g}}^{(0)}&=-0.02215(99),\qquad (\chi^{2}/\mathrm{DoF}\simeq 1.48),
	\label{eq:oneloopcoef}
\end{align} 
as the one-loop coefficient. The $\Lambda$-parameter ratio is estimated as
\begin{align}
	\frac{\Lambda_{\SF}}{c\Lambda_{\TGF}}=0.8530(61).
	\label{eq:lambda-parameter-ratio}
\end{align}


\section{The $\Lambda$-parameter in the $\MSbar$ scheme}
\label{sec:final-result}

\begin{figure}[t]
	\begin{minipage}{0.5\hsize}
		\centering
		\includegraphics[width=\figwidth\textwidth,angle=-90]{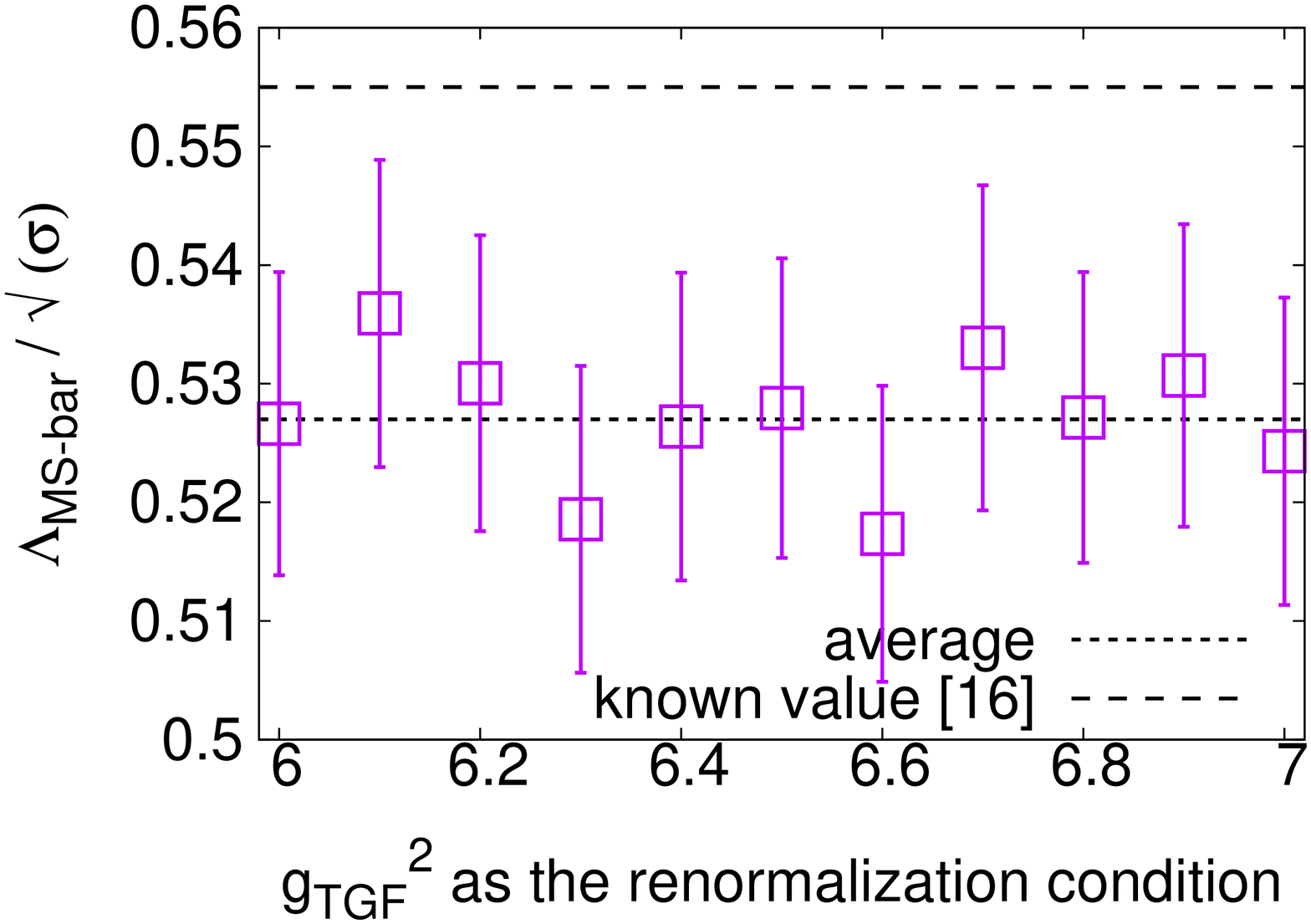}
	\end{minipage}
	\begin{minipage}{0.5\hsize}
		\centering
		\includegraphics[width=\figwidth\textwidth,angle=-90]{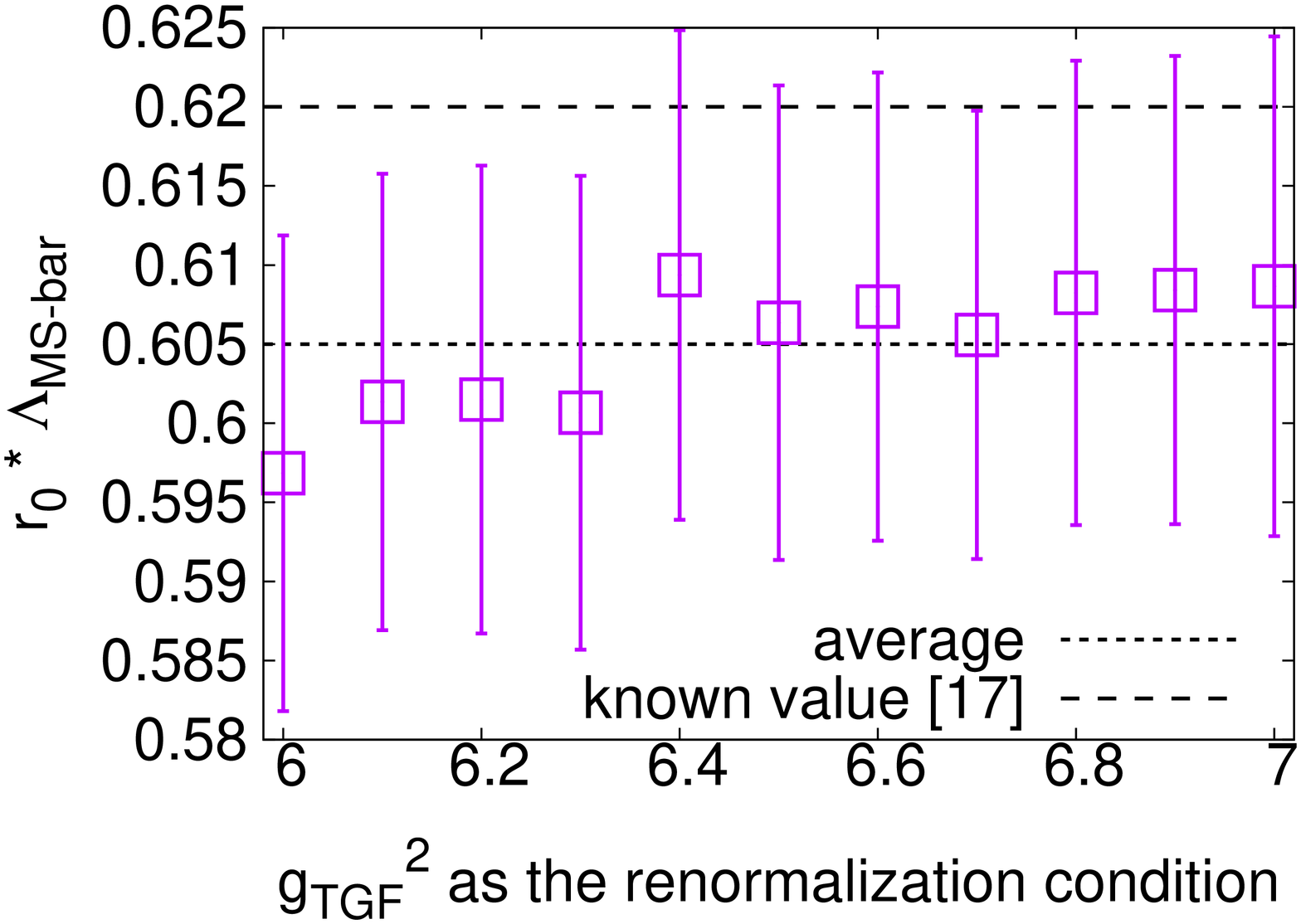}
	\end{minipage}
		\caption{The results for $\Lambda_{\MSbar}/\sqrt{\sigma}$ (left) and $r_{0}\Lambda_{\MSbar}$ (right). 	The dashed lines are average of our results. The dotted lines are the known values $\Lambda_{\MSbar}/\sqrt{\sigma}=0.555(^{+19}_{-17})$ \cite{Bali/Schilling:LambdaMSbar-StringTension} and $r_{0}\Lambda_{\MSbar}=0.62(2)$ \cite{FLAG}, respectively.}
	\label{fig:final}
\end{figure}

Substituting all pieces obtained so far into Eq. (\ref{eq:strategy}), we obtain $\Lambda_{\MSbar}/\sqrt{\sigma}$ and $r_{0}\Lambda_{\MSbar}$. The results are shown in Figure \ref{fig:final} and are independent from the choice of the initial condition on $g^2_{\TGF}(1/\Lmax)$. The statistical error from $\Lmax\Lambda_{\TGF}$ dominates the error of $\Lambda_{\MSbar}$, as seen by comparing the errors in Tables \ref{tab:LambdaTGF} and \ref{tab:string-tension.etc}, and Eq. (\ref{eq:lambda-parameter-ratio}).

From these data, we estimate the $\Lambda$-parameter in the $\MSbar$ scheme as
\begin{align}
	\Lambda_{\MSbar}/\sqrt{\sigma}=0.527(13)(10),\quad r_{0}\Lambda_{\MSbar}=0.605(15)(5).
	\label{eq:ST}
\end{align}
The first error is the statistical one and the second is the systematic one estimated from the fluctuations by the choice on $g^2_{\TGF}(1/\Lmax)$. Our results are consistent with the known values, $\Lambda_{\MSbar}/\sqrt{\sigma}=0.555(^{+19}_{-17})$ from Ref. \cite{Bali/Schilling:LambdaMSbar-StringTension} and $r_{0}\Lambda_{\MSbar}=0.62(2)$ from Ref. \cite{FLAG}, within $1.2 \sigma$ and $0.6 \sigma$ respectively.

The non-trivial part in our analysis is the use of $\Lambda_{\SF}/\Lambda_{\TGF}$ estimated from the numerical simulations on the lattice. The consistency of the $\Lambda$-parameter in the $\MSbar$ scheme strongly suggests the validity of the value for $\Lambda_{\SF}/\Lambda_{\TGF}$ in Eq. (\ref{eq:lambda-parameter-ratio}) and the one-loop expansion parameter $c_{\mathrm{g}}^{(0)}$ in Eq. (\ref{eq:oneloopcoef}). The explicit perturbative computation for the one-loop coefficient will reveal the exact value in near future \cite{bribian}.


\acknowledgments

The numerical simulations have been done on the INSAM (Institute for Nonlinear Sciences and Applied Mathematics) cluster system at Hiroshima University. This work was partly supported by JSPS KAKENHI Grant Numbers 26400249 and 16K05326.



\begin{thebibliography}{99}
	\bibitem{Ramos:TGF}
		A. Ramos, 
		\emph{JHEP} \textbf{1411} (2014) 101, 
		[arXiv:1409.1445].
	\bibitem{Narayanan:2006rf}
		R. Narayanan and H. Neuberger,
		\emph{JHEP} \textbf{0603} (2006) 064
		[hep-th/0601210].
	\bibitem{Luscher:WilsonFlow}
		M. L\"{u}scher, 
		\emph{JHEP} \textbf{1008} (2010) 071, 
		[arXiv:1006.4518].
	\bibitem{Luscher/Weisz:GF}
		M. L\"{u}scher and P. Weisz, 
		\emph{JHEP} \textbf{1102} (2011) 051, 
		[arXiv:1101.0963].
	\bibitem{Fodor/Holland/Kuti/Nogradi:pGF}
		Z. Fodor, K. Holland, J. Kuti, D. Nogradi and C.H. Wong, 
		\emph{JHEP} \textbf{1211} (2012) 007, 
		[arXiv:1208.1051].
	\bibitem{Lin:2015zpa}
		C.-J. D. Lin, K. Ogawa and A. Ramos,
		\emph{JHEP} \textbf{1512} (2015) 103
		[arXiv:1510.05755].
	\bibitem{Fritzsch/Ramos:GF}
		P. Fritzsch and A. Ramos, 
		\emph{JHEP} \textbf{1310} (2013) 008, 
		[arXiv.1301.4388].
	\bibitem{Brida:2015gqj}
		{\textbf{ALPHA}} Collaboration, M. Dalla Brida {\it et al.},
		PoS LATTICE \textbf{2015} (2016) 248
		[arXiv:1511.05831].
	\bibitem{Leino:2015bfg}
		V. Leino, T. Karavirta, J. Rantaharju, T. Rantalaiho, K. Rummukainen, J. M. Suorsa and K. Tuominen,
		PoS LATTICE \textbf{2015} (2016) 226
		[arXiv:1511.03563].
	\bibitem{bribian}
		E. Ibanez Bribian and M. Garcia Perez,
		PoS LATTICE {\textbf{2016}} 371.
	\bibitem{SFREFS}
		M. L\"{u}scher,  R. Narayanan, P. Weisz and U. Wolff,  
		\emph{Nucl. Phys.} \textbf{B384} (1992) 168-228, 
		[hep-lat/9207009],
		M. L\"{u}scher,  R. Sommer, U. Wolff and P. Weisz,  
		\emph{Nucl. Phys.} \textbf{B389} (1993) 247-264, 
		[hep-lat/9207010],
		M. L\"{u}scher,  R. Narayanan, P. Weisz and U. Wolff,
		\emph{Nucl. Phys.} \textbf{B413} (1994) 481-502, 
		[hep-lat/9309005],
		A. Bode, P. Weisz and U. Wolff, 
		\emph{Nucl. Phys.} \textbf{B576} (2000) 517-539,
		[hep-lat/9911018],
		{\textbf{ALPHA}} Collaboration, M. Della Morte et al., 
		\emph{Nucl. Phys.} \textbf{B713} (2005) 378-409,
		[hep-lat/0411025],
		{\textbf{ALPHA}} Collaboration, F. Tekin, R. Sommer and U. Wolff, 
		\emph{Nucl. Phys.} \textbf{B840} (2010) 114-128,
		[arXiv:1006.0672].
	\bibitem{Sint/Sommer:LambdaSF}
		S. Sint and R. Sommer, 
		\emph{Nucl. Phys.} \textbf{B465} (1996) 71-98,
		[hep-lat/9508012].
	\bibitem{Allton/Teper/Trivini:StringTension}
		C. Allton, M. Teper and A. Trivini, 
		\emph{JHEP} \textbf{0807} (2008) 021, 
		[arXiv:0803.1092].
	\bibitem{Antonio/Okawa:StringTension}
		A. Gonz\'{a}lez-Arroyo and M. Okawa, 
		\emph{Phys. Lett.} \textbf{B718} (2013) 1524-1528,
		[arXiv:1206.0049].
	\bibitem{Necco:Dthesis}
		S. Necco, 
		Ph.D thesis (2003)
		[hep-lat/0306005].
	\bibitem{Bali/Schilling:LambdaMSbar-StringTension}
		G. S. Bali and K. Schilling, 
		\emph{Phys. Rev.} \textbf{D47} (1993) 661-672,
		[hep-lat/9208028].
	\bibitem{FLAG}
		{\textbf{Flavour Lattice Averaging Group}}, S. Aoki et al., 
		\emph{Eur. Phys. J.} \textbf{D74} (2014) 2890,
		[arXiv:1607.00299].
\end{thebibliography}
\end{document}